\documentclass[
  journal=pasa,
  manuscript=research-paper, 
  year=202X,
  volume=YY,
]{cup-journal}

\usepackage{amsmath}
\usepackage{amssymb,microtype,siunitx,booktabs}
\DeclareUnicodeCharacter{02BC}{'}
\usepackage[colorlinks=true, linkcolor=blue, citecolor=blue, urlcolor=blue]{hyperref}
\usepackage{xcolor}

\title{Radio frequency interference identification using eigenvalue decomposition for multi-beam observations}

\email[Shi Dai, Na Wang]{Shi.Dai@csiro.au, na.wang@xao.ac.cn}

\author{Juntao Bai} 
\affiliation{Xinjiang Astronomical Observatory, Chinese Academy of Sciences, Urumqi, Xinjiang 830011, Peopleʼs Republic of China}
\alsoaffiliation{Institute for Gravitational Wave Astronomy, Henan Academy of Sciences, Zhengzhou 450046, Henan, People’s Republic of China}

\author{Shi Dai} 
\affiliation{Australia Telescope National Facility, CSIRO, Space and Astronomy, PO Box 76, Epping, NSW 1710, Australia}
\alsoaffiliation{Western Sydney University, Locked Bag 1797, Penrith South DC, NSW 2751, Australia}

\author{Na Wang} 
\affiliation{Xinjiang Astronomical Observatory, Chinese Academy of Sciences, Urumqi, Xinjiang 830011, Peopleʼs Republic of China}
\alsoaffiliation{State Key Laboratory of Radio Astronomy and Technology, A20 Datun Road, Chaoyang District, Beijing, 100101, Peopleʼs Republic of China}
\alsoaffiliation{Xinjiang Key Laboratory of Radio Astrophysics, 150 Science1-Street, Urumqi, Xinjiang 830011, Peopleʼs Republic of China}

\author{Stefan Os{\l}owski} 
\affiliation{Australia Telescope National Facility, CSIRO, Space and Astronomy, PO Box 76, Epping, NSW 1710, Australia}

\author{Shuangqiang Wang} 
\affiliation{Xinjiang Astronomical Observatory, Chinese Academy of Sciences, Urumqi, Xinjiang 830011, Peopleʼs Republic of China}
\alsoaffiliation{State Key Laboratory of Radio Astronomy and Technology, A20 Datun Road, Chaoyang District, Beijing, 100101, Peopleʼs Republic of China}
\alsoaffiliation{Xinjiang Key Laboratory of Radio Astrophysics, 150 Science1-Street, Urumqi, Xinjiang 830011, Peopleʼs Republic of China}
\alsoaffiliation{Australia Telescope National Facility, CSIRO, Space and Astronomy, PO Box 76, Epping, NSW 1710, Australia}

\author{George Hobbs} 
\affiliation{Australia Telescope National Facility, CSIRO, Space and Astronomy, PO Box 76, Epping, NSW 1710, Australia}

\author{Jianping Yuan} 
\affiliation{Xinjiang Astronomical Observatory, Chinese Academy of Sciences, Urumqi, Xinjiang 830011, Peopleʼs Republic of China}
\alsoaffiliation{State Key Laboratory of Radio Astronomy and Technology, A20 Datun Road, Chaoyang District, Beijing, 100101, Peopleʼs Republic of China}
\alsoaffiliation{Xinjiang Key Laboratory of Radio Astrophysics, 150 Science1-Street, Urumqi, Xinjiang 830011, Peopleʼs Republic of China}

\author{Wenming Yan} 
\affiliation{Xinjiang Astronomical Observatory, Chinese Academy of Sciences, Urumqi, Xinjiang 830011, Peopleʼs Republic of China}
\alsoaffiliation{State Key Laboratory of Radio Astronomy and Technology, A20 Datun Road, Chaoyang District, Beijing, 100101, Peopleʼs Republic of China}
\alsoaffiliation{Xinjiang Key Laboratory of Radio Astrophysics, 150 Science1-Street, Urumqi, Xinjiang 830011, Peopleʼs Republic of China}

\author{Qijun Zhi} 
\affiliation{School of Physics and Electronic Science, Guizhou Normal University, Guiyang 550001, Peopleʼs Republic of China}
\alsoaffiliation{Guizhou Provincial Key Laboratory of Radio Astronomy and Data Processing, Guizhou Normal University, Guiyang 550001, Peopleʼs Republic of China.}

\author{Lunhua Shang} 
\affiliation{School of Physics and Electronic Science, Guizhou Normal University, Guiyang 550001, Peopleʼs Republic of China}
\alsoaffiliation{Guizhou Provincial Key Laboratory of Radio Astronomy and Data Processing, Guizhou Normal University, Guiyang 550001, Peopleʼs Republic of China.}

\author{Xin Xu} 
\affiliation{School of Physics and Electronic Science, Guizhou Normal University, Guiyang 550001, Peopleʼs Republic of China}
\alsoaffiliation{Guizhou Provincial Key Laboratory of Radio Astronomy and Data Processing, Guizhou Normal University, Guiyang 550001, Peopleʼs Republic of China.}

\author{Shijun Dang} 
\affiliation{School of Physics and Electronic Science, Guizhou Normal University, Guiyang 550001, Peopleʼs Republic of China}
\alsoaffiliation{Guizhou Provincial Key Laboratory of Radio Astronomy and Data Processing, Guizhou Normal University, Guiyang 550001, Peopleʼs Republic of China.}

\author{De Zhao} 
\affiliation{Xinjiang Astronomical Observatory, Chinese Academy of Sciences, Urumqi, Xinjiang 830011, Peopleʼs Republic of China}
\alsoaffiliation{State Key Laboratory of Radio Astronomy and Technology, A20 Datun Road, Chaoyang District, Beijing, 100101, Peopleʼs Republic of China}
\alsoaffiliation{Xinjiang Key Laboratory of Radio Astrophysics, 150 Science1-Street, Urumqi, Xinjiang 830011, Peopleʼs Republic of China}

\received {dd Mmm YYYY}
\revised  {dd Mmm YYYY}
\accepted {dd Mmm YYYY}
\published{22 September 202X}


\begin{document}

\begin{abstract}
With the installation of next-generation phased array feed (PAF) receivers on radio telescopes, there is an urgent need to develop effective and computationally efficient radio frequency interference (RFI) mitigation methods for large-scale surveys. Here we present a new RFI mitigation package, called \texttt{mRAID} (multi-beam RAdio frequency Interference Detector), which uses the eigenvalue decomposition algorithm to identify RFI in cross-correlation matrix (CCM) of data recorded by multiple beams. 
When applied to high time-resolution pulsar search data from the Five-hundred-meter Aperture Spherical Radio Telescope (FAST), \texttt{mRAID} demonstrates excellent performance in identifying RFI over short timescales, thereby enhancing the efficiency of pulsar and fast radio burst (FRB) searches. Since the computation of the CCM and the eigenvalue decomposition for each time sub-integration and frequency channel are independent, the process is fully parallelisable. As a result, \texttt{mRAID} offers a significant computational advantage over commonly used RFI detection methods.
\end{abstract}

\section{Introduction} \label{sec:intro}

Radio Frequency Interference (RFI) has become an increasingly severe challenge for radio astronomy, driven by the growing sensitivity of modern telescopes and the proliferation of human-made radio sources. 
Time-domain sciences like pulsar and fast radio burst (FRB) searches often rely on high time-resolution observations, which can be significantly impacted by both impulsive and persistent RFI. Moreover, identifying and masking RFI in such observations can demand substantial computational resources. Consequently, the development of effective and efficient RFI mitigation algorithms is crucial for maximizing the scientific outputs of astronomical data.

Over the years, numerous RFI mitigation strategies have been proposed, including frequency-domain approaches such as the use of notch filters, time-domain techniques such as data flagging, and telescope-level measures such as the establishment of radio-quiet zones \citep{Fridman2001A&A, Briggs2005astro, Gary2010PASP, Baan2019JAI}. Among these, threshold-based methods are the most widely adopted due to their simplicity and efficiency \citep[e.g.,][]{Ransom2002AJ, Offringa2010MNRAS, Zeng2021MNRAS}. 
More advanced techniques, such as median filtering and SumThreshold algorithms, have also been developed to enhance RFI detection \citep{Bhat2005RaSc, Offringa2010MNRAS, Zeng2021MNRAS}. However, these methods often struggle with weak and persistent RFI and are limited by fixed time-frequency resolution settings \citep{Baan2004AJ}.

Machine learning approaches represent a promising new direction in RFI mitigation. Techniques such as convolutional neural networks (CNNs) and U-Net architectures have been employed to automatically learn and classify RFI features \citep{Akeret2017A&C, Burd2018AN, Kerrigan2019MNRAS, Vafaei2020MNRAS, Dao2024A&C}. While these methods show great potential, they require large, labelled datasets for training, making them time-consuming and challenging to implement for large-scale observatories.

Radio observations conducted with multi-beam receivers or phased array feeds (PAF) offer a significant advantage for mitigating RFI, as most RFI signals are typically detected by multiple beams or pixels.
In high time-resolution observations using multi-beam systems, spatial filtering techniques were explored by \citet{Kocz2010AJ} for the 13-beam receiver on Murriyang, the Parkes 64-metre radio telescope. Similar approaches have since been implemented for the 19-beam receiver system on the Five-hundred-meter Aperture Spherical radio Telescope (FAST) \citep{Wang2022A&C}. While spatial filtering has proven to be effective \citep{Kocz2010AJ}, it relies on the covariance matrix of multi-beam voltage data, which is generally unavailable for large-scale pulsar and FRB surveys \footnote{See, e.g., \url{https://www.atnf.csiro.au/observers/data/ppdu_guide.html} and \url{https://fast.bao.ac.cn/cms/article/129/}.}.
As an alternative, the cross-correlation matrix (CCM) of multi-beam power data can be utilized to identify and mitigate RFI. For instance, \citet{kbb+12} applied singular value decomposition (SVD) to the CCM of frequency-averaged (at zero dispersion measure) time series from Murriyang’s multi-beam pulsar surveys, significantly reducing the number of false candidates. More recently, \citet{Chen2023RAA} developed an RFI mitigation pipeline for FAST’s 19-beam surveys, employing the SumThreshold algorithm on the CCM to effectively manage RFI.

In this study, we expand upon the method developed by \citet{kbb+12}, extending it to both the time and frequency domains to enhance the efficiency of RFI identification for multi-beam receivers and future PAF systems. The CCM is computed for each frequency channel and a specified integration time, and eigenvalue decomposition is applied to identify RFI across different beams. Additionally, we incorporate the Asymmetric Reweighted Penalized Least Squares (ArPLS) algorithm \citep{Baek2015Ana} to fit and normalise bandpass, improving sensitivity to weak RFI. A publicly available package, called mRAID (multi-beam RAdio frequency Interference Detector), is provided alongside this paper. 
Section \ref{sec:2} outlines the data set used in this study. In Section \ref{sec:3}, we describe the key stages of our RFI identification methods, including bandpass normalisation, the construction of the CCM, and the application of the eigenvalue decomposition method. Section \ref{sec:4} details the implementation of the proposed approach on FAST data and discusses the results. Section \ref{sec:5} gives some conclusions.

\section{Experimental data} \label{sec:2}

To demonstrate the performance of \texttt{mRAID}, we used one observation from the FAST \footnote{\url{http://fast.bao.ac.cn}} 19-beam receiver in the pulsar search mode, where PSR~J1832+0204\_P was discovered in beam 11~\citep{Han2025RAA}.
The observation has a time resolution of 49.152\,$\mu$s and a frequency resolution of 122.07\,kHz, spanning the range of 1000 to 1500 MHz \citep{Jiang2020RAA}. The total integration time is 412.00\,s, and there are a total of 8192 sub-integrations in the file. Each sub-integration consists of $1024\times4096$ data points, where 1024 represents the number of sampling points per sub-integration, and 4096 corresponds to the number of frequency channels. The data was 8-bit sampled and two polarisations were recorded.

For the FAST 19-beam receiver, satellite and aircraft signals are the two primary sources of RFI. Satellite RFI typically occupies multiple frequency bands, with each band often shared by several satellites. These interferences are broadband and persist throughout the observation. In contrast, aircraft RFI contaminates only a few pixels in the time-frequency plane and lasts for a few seconds. Besides these known RFI sources, the data often include other narrowband, continuous, or intermittent interference signals.

\section{RFI identification strategy} \label{sec:3}

Our RFI identification scheme is designed based on three key components: the bandpass normalisation, the construction of the CCM, and the identification of RFI using eigenvalue decomposition. We describe each of these components in the following sections.

\subsection{Bandpass normalisation using ArPLS}
\label{sec:3.1}

The signal level can vary considerably across different beams and frequencies, which can affect the performance of CCM-based RFI identification.  To improve the robustness of the method, especially for identifying low-level RFI, bandpass normalisation is therefore highly beneficial.
The ArPLS algorithm is a robust method tailored for this purpose, offering significant advantages over traditional approaches such as polynomial fitting or Gaussian filtering. By iteratively minimising a regularized least-squares function, ArPLS balances the need for accurately fitting the underlying signal and suppressing noise or RFI contamination. As demonstrated by previous studies~\citep[e.g.,][]{Zeng2021MNRAS, Wang2022A&C}, ArPLS outperforms conventional methods in both speed and accuracy, making it an ideal choice for large-scale surveys that require efficient bandpass normalisation as part of the RFI mitigation process.

To perform bandpass normalisation, we first aggregate the time-domain samples within a given time interval (e.g., 1.00\,s) for each frequency channel. The bandpass shape is then fit using the ArPLS method. During normalisation, the fitted bandpass is subtracted from the time-domain data for each frequency channel. The correction is applied uniformly across all time samples, ensuring that the bandpass variations are consistently removed without altering the original resolution of the data.

\begin{figure} [!htp]
   \centering
   \includegraphics[width=8cm]{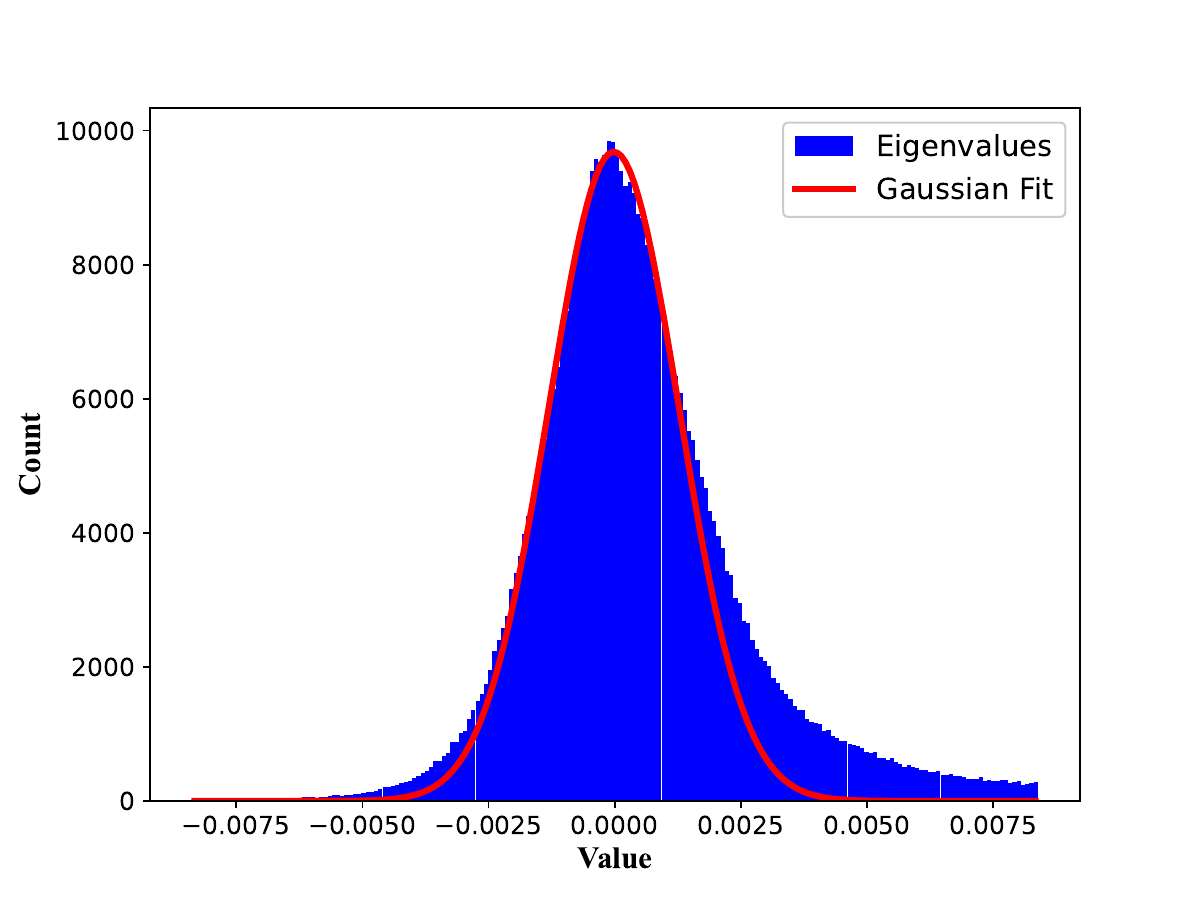}
   \caption{Histogram of dominant eigenvalues derived for all subintervals and channels of the FAST 19-beam test observation. A Gaussian fit to the distribution is shown as the red line.}
   \label{Eigval_fit} 
\end{figure}

\begin{figure*} [!htp]
   \centering
   \includegraphics[width=8cm]{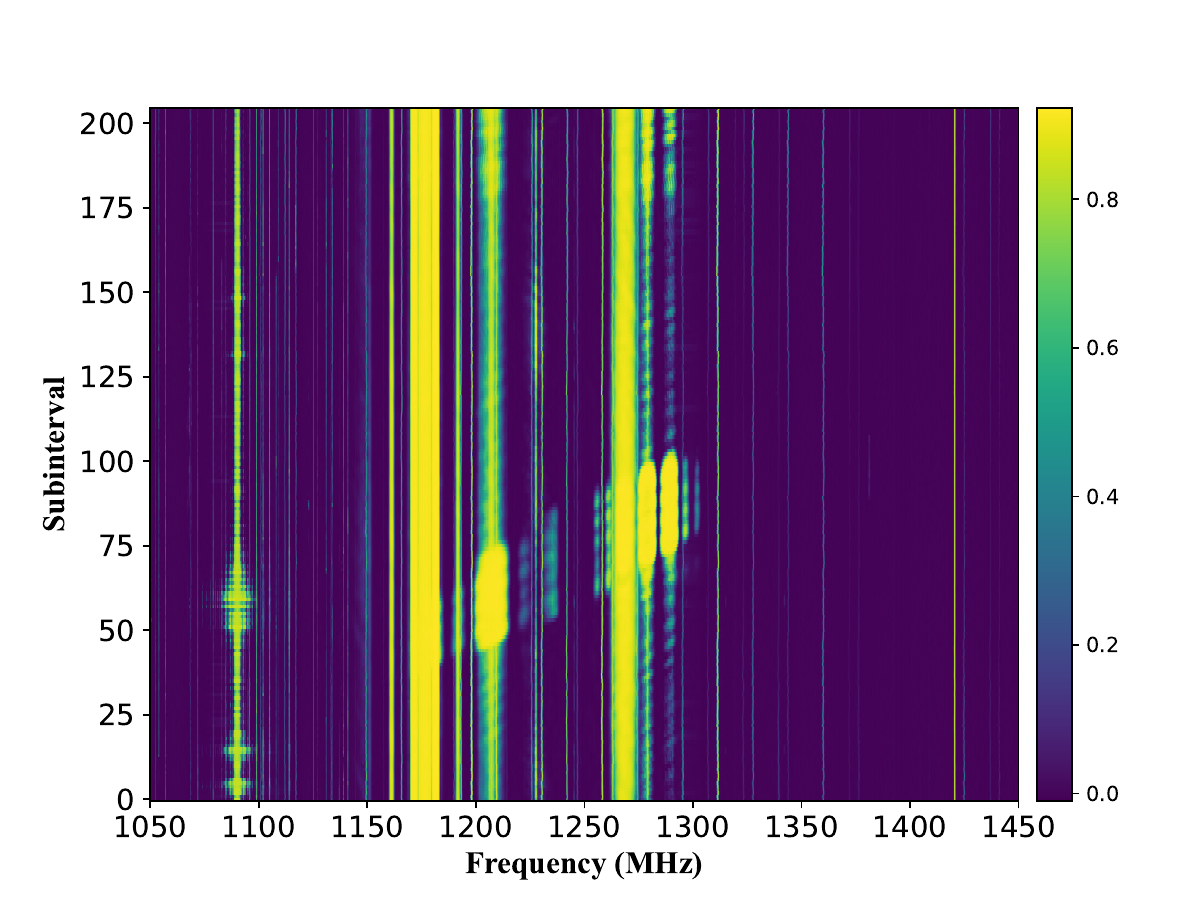}
   \includegraphics[width=8cm]{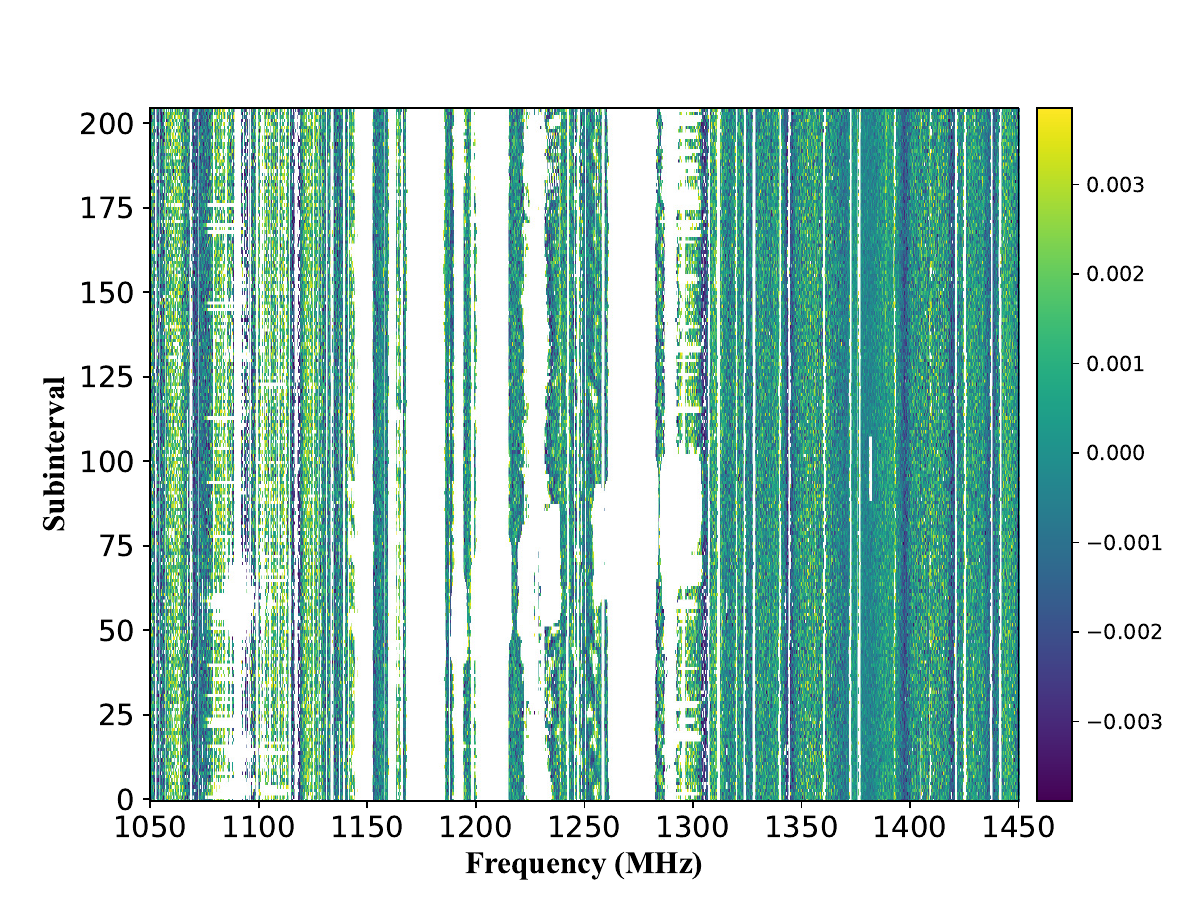}
   \caption{The left panel shows dominant eigenvalues derived for the FAST 19-beam test observation. The right panel shows the result after masking out RFI affected subintervals and channels, where the eigenvalue distribution becomes noise-like, indicating the effectiveness of the iterative RFI identification process.}
   \label{Eigval} 
\end{figure*}

\subsection{The construction of CCM}
\label{sec:3.2}

After bandpass normalisation, we calculate the CCM for a given frequency channel ($j$) and subinterval ($k$), which is defined as
\begin{equation}
C_{\rm j,k} = 
\begin{pmatrix}
\langle B_{\rm j,k,1} B_{\rm j,k,1} \rangle & \langle B_{\rm j,k,1} B_{\rm j,k,2} \rangle & \cdots & \langle B_{\rm j,k,1} B_{\rm j,k,i} \rangle \\
\langle B_{\rm j,k,2} B_{\rm j,k,1} \rangle & \langle B_{\rm j,k,2} B_{\rm j,k,2} \rangle & \cdots & \langle B_{\rm j,k,2} B_{\rm j,k,i} \rangle \\
\vdots & \vdots & \ddots & \vdots \\
\langle B_{\rm j,k,i} B_{\rm j,k,1} \rangle & \langle B_{\rm j,k,i} B_{\rm j,k,2} \rangle & \cdots & \langle B_{\rm j,k,i} B_{\rm j,k,i} \rangle
\end{pmatrix}
\end{equation}
Here, $B_{\rm j,k,i}$ represents the total power recorded from the $i$-th beam in channel $j$ and subinterval $k$, which is a function of time. The number of time sample in $B_{\rm j,k,i}$ is $N_{\rm s}=T_{\rm k}/t_{\rm s}$, where $T_{\rm k}$ is the length of subinterval and $t_{\rm s}$ is the sampling time. Each element in the matrix, denoted as $\langle \cdots \rangle$, represents the cross-correlation between two beams for a given subinterval $k$ and channel $j$.
The correlation matrix $C_{\rm j,k}$ is a square $N_{\rm beam} \times N_{\rm beam}$ matrix.

In multi-beam receiver systems, significant RFI often creates strong correlations between signals from different beams. Calculating the CCM provides an efficient way for identifying RFI while also improving sensitivity to low-level interference compared to algorithms that analyse each beam independently.
By performing eigenvalue decomposition on the cross-correlation matrix of the multi-beam data, we can isolate these common structures, which are likely dominated by RFI.

\subsection{The identification of RFI using eigenvalue decomposition}

Eigenvalue decomposition is a matrix factorisation technique that decomposes a square $n\times n$ matrix into:
\begin{equation}
    C = Q \Lambda Q^{-1},
\end{equation}
where $Q$ is a square $n\times n$ matrix whose $i$th column is the eigenvector $q_{i}$ of $C$, and $\Lambda$ is a diagonal matrix whose diagonal elements are the corresponding eigenvalues, $\Lambda_{\rm i,i}=\lambda_{\rm i}$. These eigenvalues are normalised by squaring them and dividing by the sum of all squared eigenvalues, ensuring that the eigenvalues reflect the proportion of variance explained by each eigenvector.
The eigenvalues in $\Lambda$ are ordered by magnitude, from the largest to the smallest ($\lambda _1 > \lambda _2 > \dots > \lambda _i$). Large eigenvalues correspond to the dominant features in the CCM, which are often attributed to RFI affecting multiple beams. The eigenvectors associated with these large eigenvalues allow us to identify beams most affected by these RFI.

\begin{figure} [!htp]
   \centering
   \includegraphics[width=8cm]{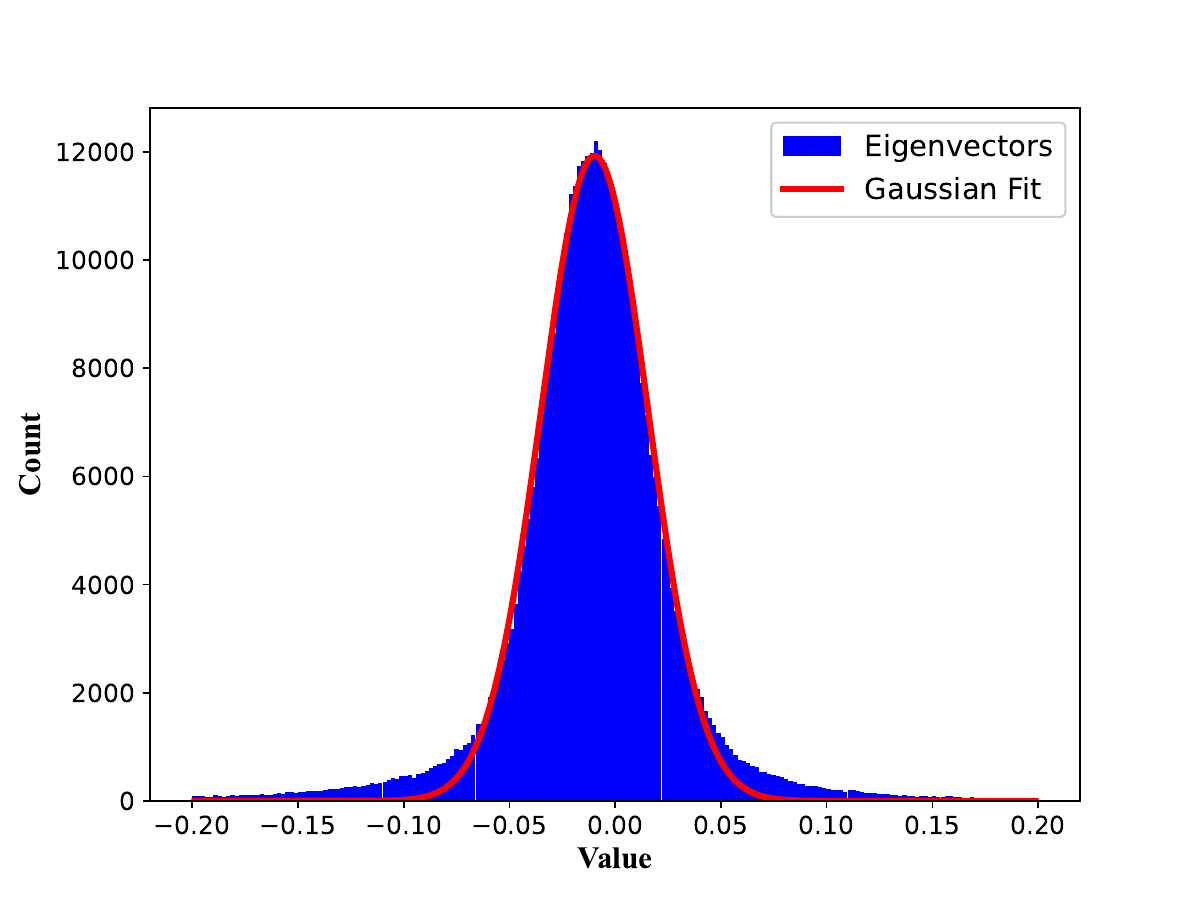}
   \caption{Histogram of dominant eigenvectors of non-RFI channels (those not flagged as RFI in Figure \ref{Eigval_fit}) derived for the FAST 19-beam test observations. A Gaussian fit to the distribution is shown as the red line.}
   \label{Eigvec_fit} 
\end{figure}

For a given subinterval and channel of an observation, \texttt{mRAID} calculates its CCM ($C_{\rm j,k}$) and performs eigenvalue decomposition. The largest eigenvalue and its associated eigenvectors are stored in a HDF5 \footnote{https://www.hdfgroup.org/} format file, to be further processed after the whole observation is completed. Such a process is completely independent for different subintervals and channels and can be carried out in parallel. 
Once the eigenvalue decomposition is finished for all subintervals, RFI-contaminated channels are identified through an iterative process that masks outliers in eigenvalues as a function of frequency and time (i.e. subinterval). The mean and standard deviation of eigenvalues are measured by performing a Gaussian fit to the distribution (see Figure \ref{Eigval_fit}), and iteratively masks channels with values exceeding a dynamic threshold. Once convergence is reached, a combined RFI mask is generated, capturing all identified RFI-contaminated channels and subintervals (see Figure \ref{Eigval}).

After masking eigenvalues, the next step involves using their associated eigenvectors to further confirm whether identified RFI channels are present across all beams.
By applying a threshold to the elements of these vectors, the RFI detection process can be refined. If the magnitude of an element exceeds the threshold, the corresponding beam is considered to be strongly affected by RFI. Conversely, elements below the threshold indicate that the beams are unaffected and are classified as non-RFI channels. The threshold for the eigenvectors is determined individually for each beam by fitting a Gaussian distribution to the eigenvector elements of non-RFI channels specific to that beam as shown in Figure \ref{Eigvec_fit}. This beam-specific approach ensures a more accurate classification of RFI channels tailored to the unique characteristics of each beam.

\section{Experimental results and discussion} \label{sec:4}

To evaluate the effectiveness of \texttt{mRAID} in identifying RFI, we performed a comparative analysis using observational data from the FAST 19-beam receiver. For comparison, we used \texttt{rfifind}, the RFI masking tool included in the widely used pulsar search software package \textit{PRESTO}. Although several new tools have been developed specifically for FAST~\citep[e.g.,][]{Zeng2021MNRAS, Wang2022A&C, Chen2023RAA}, threshold-based methods, such as \texttt{rfifind}, remain the standard RFI mitigation tool for major FAST pulsar surveys~\citep[e.g.,][]{Li2018IMMag,Qian2019SCPMA, Han2021RAA}. As such, \texttt{rfifind} serves as an appropriate benchmark for assessing \texttt{mRAID}'s performance.

\begin{figure*} [!htp]
   \centering
   \includegraphics[width=16cm]{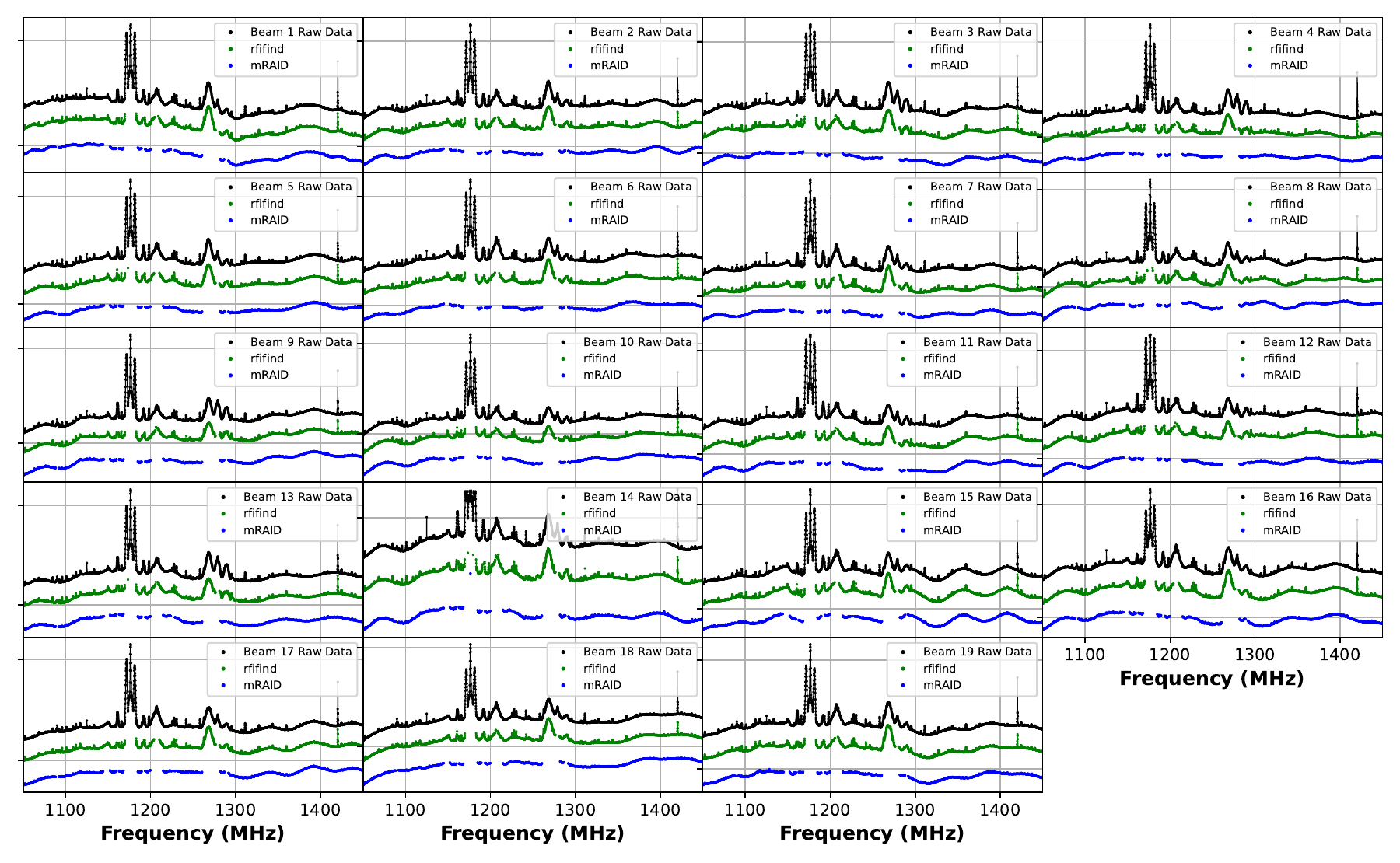}
   \caption{Time averaged spectrum of each beam of the FAST observation. Raw spectra are shown as black points; results of \texttt{rfifind} are shown as green; results of \texttt{mRAID} are shown as blue. Compared with the raw spectra, while both methods effectively identify strongly RFI-affected channels, \texttt{mRAID} shows superior performance in identifying weak RFI. } 
   \label{all_beam_vs} 
\end{figure*}

\begin{figure*} [!htp]
   \centering
   \includegraphics[width=8cm]{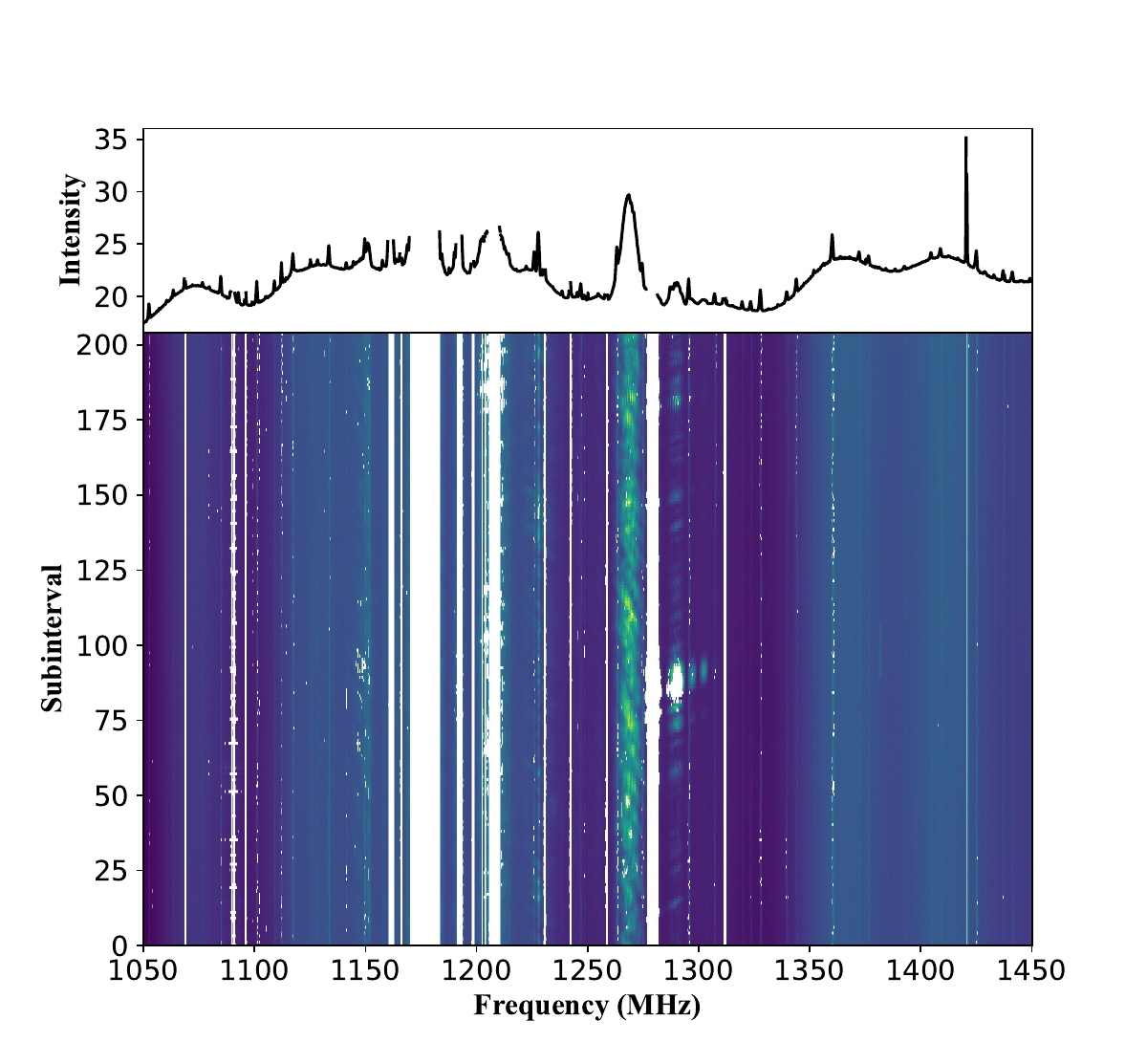}
   \includegraphics[width=8cm]{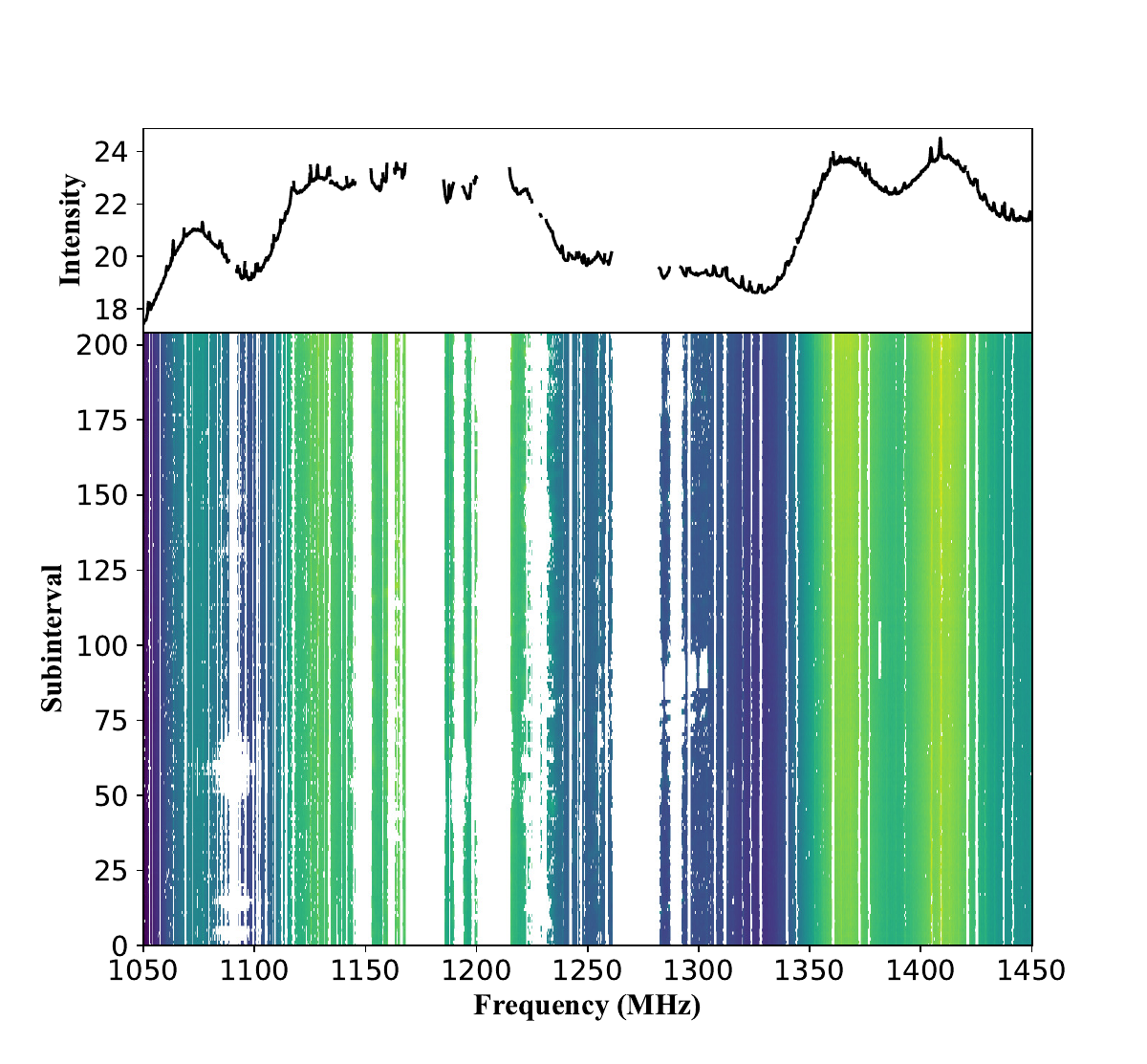}
   \caption{RFI masked data of beam 11 of the FAST observation. Left: results of \texttt{rfifind}; right: results of \texttt{mRAID}. Compared with \texttt{rfifind}, \texttt{mRAID} produces much cleaner data in time and frequency after masking, effectively preserving uncontaminated regions while removing both narrow-band and broadband RFI. }
   \label{M11_vs} 
\end{figure*}

\subsection{Frequency domain RFI}

In Figure \ref{all_beam_vs}, we compare the bandpass of FAST observations before and after RFI masking. Each subplot corresponds to a single beam, displaying the time averaged bandpass of raw data (black points), RFI-masked data using \texttt{rfifind} (green points), and RFI-masked data using \texttt{mRAID} (blue points). The horizontal axis represents frequency channels, while the vertical axis indicates signal intensity on a log scale. Strong RFI contamination is evident in the raw data, characterised by sharp peaks at specific channels. For both \texttt{rfifind} and \texttt{mRAID}, the length of subintervals is 2.00\,s.
While \texttt{rfifind} identified strong RFI, residual contamination remains, particularly in highly affected regions. In contrast, \texttt{mRAID} achieved more effective RFI suppression, yielding cleaner bandpass.

Figure \ref{M11_vs} compares RFI masked data using \texttt{rfifind} (left panel) with that using \texttt{mRAID} (right panel) for beam 11, where PSR~J1832+0204\_P was discovered. The horizontal axis represents frequency channels, while the vertical axis denotes subinterval indices. Colour regions correspond to non-RFI data, whereas white regions indicate flagged RFI. The top panel in each subplot shows the average spectrum of the masked data. \texttt{rfifind} generated a less effective RFI mask, leaving some contaminated regions insufficiently flagged,  for both narrowband (for example at $\sim1360$\,MHz) and broadband (at $\sim1270$\,MHz) RFI. 
In contrast, \texttt{mRAID} exhibited a more effective RFI suppression in both narrowband and broadband cases, identifying and excising most interference.

\begin{figure*} [!htp]
   \centering
   \includegraphics[width=8cm]{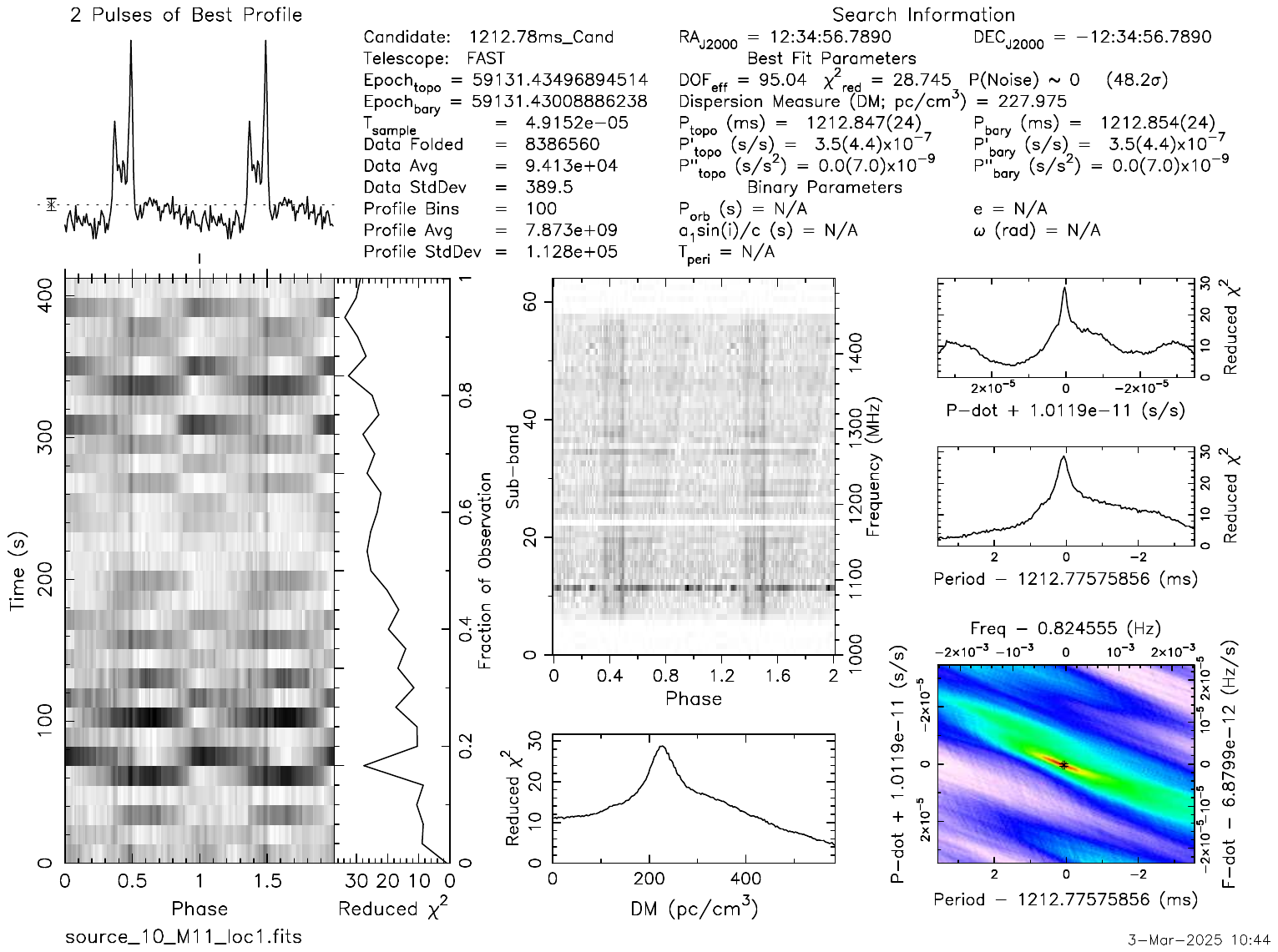}
   \includegraphics[width=8cm]{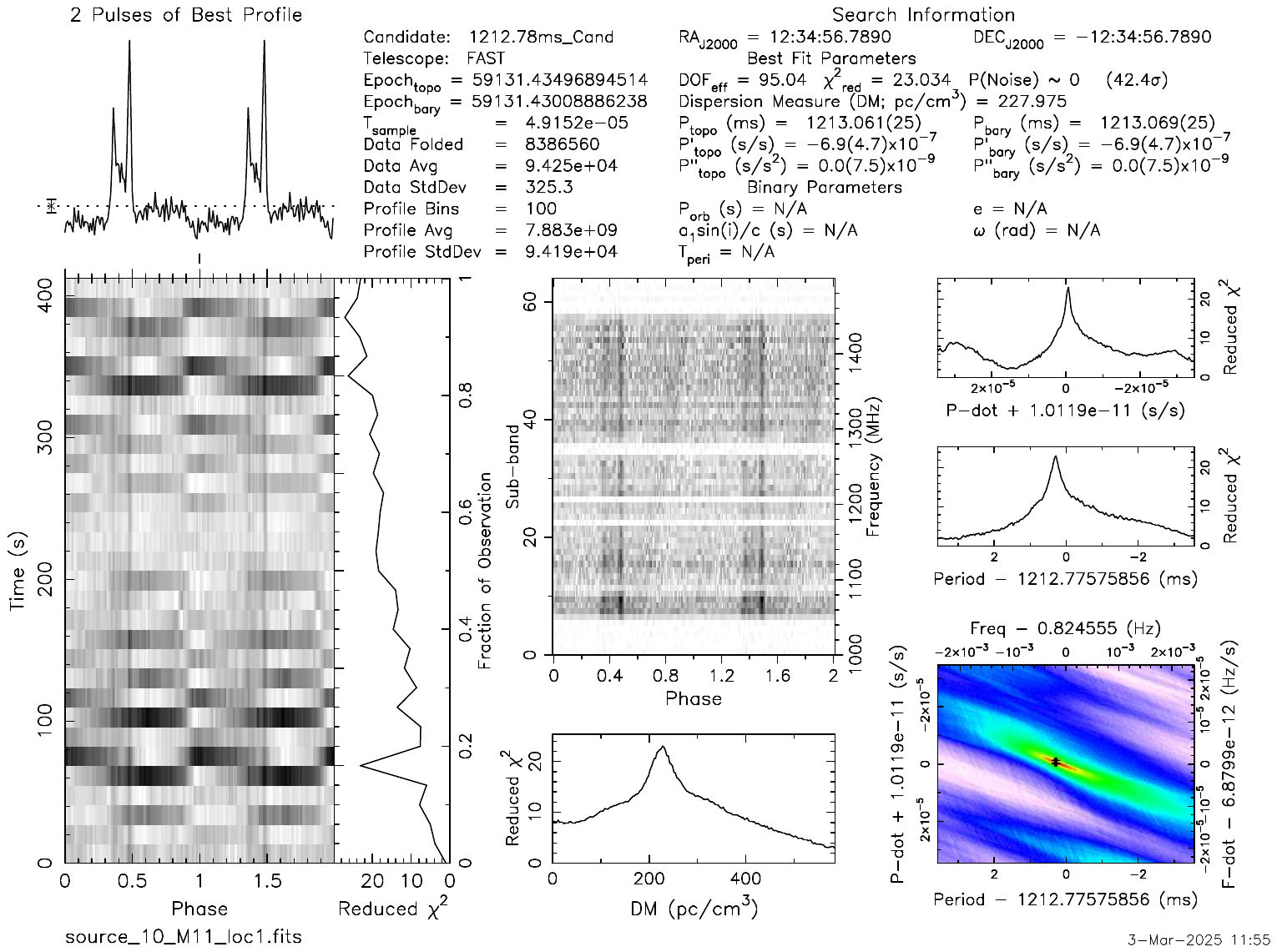}
   \caption{Comparison of the folded pulsar results after RFI masking using the \texttt{rfifind} (left) and \texttt{mRAID} (right). These plots were generated using the \texttt{prepfold} command as part of \textit{PRESTO}. \texttt{mRAID} performs a more thorough removal of weak narrow-band RFI than \texttt{rfifind}. }
   \label{fold_vs} 
\end{figure*}

To validate the algorithm's reliability, we applied the RFI mask generated by \texttt{mRAID} and folded the data with parameters of PSR~J1832+0204\_P. Figure \ref{fold_vs} presents a comparison of folded pulsar data after RFI masking using \texttt{rfifind} (left panel) and \texttt{mRAID} (right panel). Each panel includes key diagnostic plots from the pulsar folding process: the pulse profile (top), time-phase diagram (left lower), frequency-phase diagram (center), dispersion measure (DM) curve, and reduced $\chi^2$ plot. While the pulsar can be clearly detected in both cases, \texttt{mRAID} outperformed \texttt{rfifind}, resulting in a cleaner frequency-phase diagram.

\subsection{Time domain RFI}

\begin{figure*} [!htp]
   \centering
   \includegraphics[width=16cm]{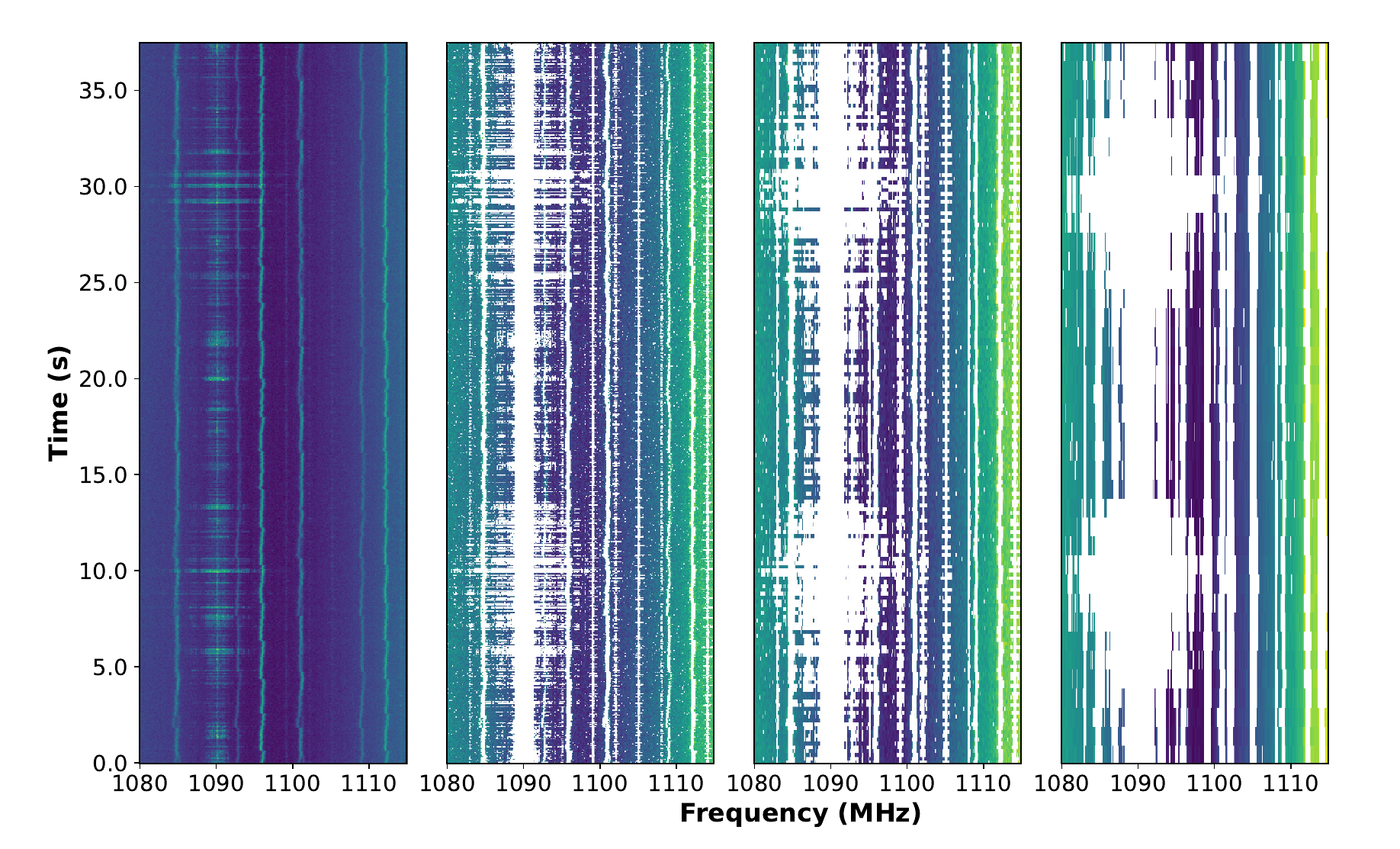}
   \caption{RFI masking results of \texttt{mRAID} at different time resolutions. The leftmost panel shows the original dynamic spectrum (unmasked), while the second to fourth panels present the results after applying \texttt{mRAID} with increasing subinterval lengths of 0.05\,s, 0.20\,s and 1.00\,s, respectively. The data correspond to the same frequency subband (1080-1115 MHz). These plots demonstrate that \texttt{mRAID} excels at identifying time-domain RFI with sub-second durations and/or periodic patterns.}
   \label{fig7} 
\end{figure*}

The CCM of multi-beam data offers an efficient approach for identifying RFI on much shorter timescales compared to methods that analyse each beam independently.
Figure \ref{fig7} demonstrates the performance of \texttt{mRAID} in identifying and masking RFI at different time resolutions. The leftmost panel shows the original unmasked dynamic spectrum, while the second to fourth panels display the results after applying \texttt{mRAID} with subinterval lengths of 0.05\,s, 0.20\,s and 1.00\,s, respectively. All panels correspond to the same frequency subband (1080-1115 MHz), with time increasing along the vertical axis and frequency along the horizontal axis.
At the finest time resolution (0.05\,s), transient and narrowband RFI structures are clearly resolved, and \texttt{mRAID} effectively masks these features. As the subinterval time increases, RFI signals are increasingly smeared in time and become more dominant in the averaged spectra. Nonetheless, \texttt{mRAID} maintains robust suppression performance across time scales, successfully identifying both weak and strong RFI features without introducing excessive masking. These results highlight the algorithm’s adaptability to varying temporal resolutions and its advantage over static, threshold-based methods in dynamic RFI environments.

We compare \texttt{mRAID} to \texttt{rfifind} at different time resolutions by applying both to the same data set. Table~\ref{table1} summarises the RFI masking rates and the number of periodic and transient candidates detected after search. For \texttt{rfifind}, the RFI masking rate gradually increases from 2.754\% at 0.05\,s to 26.608\% at 10.00\,s, and the number of periodic and transient candidates across all DM ranges decreases. This outcome is expected, as threshold-based methods are more effective at detecting weak RFI when the integration time is longer. However, for short-duration, transient RFI, these methods tend to be less effective.
In contrast, \texttt{mRAID} exhibits a significantly more effective RFI masking rate at all timescales. At 0.05\,s, \texttt{mRAID} masks 21.523\% of the data, nearly eight times that of \texttt{rfifind} at the same resolution. As the integration time increases to 2.00\,s, the masking rate of \texttt{mRAID} reaches 34.654\%, compared to 10.288\% for \texttt{rfifind} at the same time resolution. This more effective RFI mitigation strategy leads to a smaller number of surviving candidates. For example, at 2.00\,s and in the 100--200\,cm$^{-3}$\,pc DM range, \texttt{mRAID} yields 98 candidates, while \texttt{rfifind} yields 143. 
Similarly, the number of transient candidates detected, with signal-to-noise ratios (S/N) $\geq 7$, is consistently lower for \texttt{mRAID} compared to \texttt{rfifind} across all DM ranges and time resolutions. For instance, at 2.00\,s, \texttt{rfifind} detects 33969 candidates at DM 0--100\,cm$^{-3}$\,pc, whereas \texttt{mRAID} detects only 3283. This significant reduction suggests that \texttt{mRAID} suppresses false transient detection more effectively, further improving our sensitivity to weak bursts.
Our results indicate that \texttt{mRAID} provides more efficient and effective RFI masking than \texttt{rfifind}, particularly in environments with short-duration ($<1.00$\,s) interference. With significantly reduced number of candidates, \texttt{mRAID} improves the likelihood of detecting radio bursts by minimising false positives induced by RFI.

\begin{table*}
\centering
\caption{ The average percentage of RFI masks for FAST 19 beams data using the \texttt{rfifind} and \texttt{mRAID} methods, as well as the number of the periodic and transient candidates generated using the mask files within different DM ranges.}
\begin{tabular}{ccccccccccc}
\hline
\hline
Methods   &    Time    & RFI masking  &    \multicolumn{8}{c}{Number of the candidates }    \\ 
          & resolution &    rate     &    \multicolumn{5}{c}{Periodic signals (S/N $\geq$ 10)}   &    \multicolumn{3}{c}{ Transient bursts (S/N $\geq$ 7)}    \\ 
          &  (s)      &     (\%)  & \multicolumn{5}{c}{DM range (cm$^{-3}$ pc)}  & \multicolumn{3}{c}{DM range (cm$^{-3}$ pc) }  \\ 
          &     &        & {0--100}  & {100--200}  & {200--300} & {300--400}  & {400--500} & {0--100}  & {100--200}  & {200--300}   \\ 
\hline
\texttt{rfifind}  & 0.05    &     2.754    &  277  & 170 &  128    & 111 &  56  & 71484 & 3864 & 349   \\
              &     0.20     &     3.122    &  305  & 173 &  129    & 115 &  50  & 69294 & 3537 & 337   \\
              &     0.50     &     4.863    &  305  & 165 &  131    & 103 &  48  & 66783 & 3696 & 319   \\
              &     1.00     &     7.238    &  304  & 160 &  124    & 100 &  43  & 56337 & 2718 & 273   \\
              &     2.00     &    10.288    &  273  & 143 &  119    & 96  &  40  & 33969 & 1353 & 136   \\
              &     5.00     &    17.850    &  131  & 121 &  104    & 91  &  26  & 10569 & 320  & 24   \\
              &     10.00    &    26.608    &  114  & 109 &  102    & 92  &  29  & 7658  & 225  & 10    \\
\texttt{mRAID}  &   0.05    &   21.523    &  149  & 120 &  115  & 89  &  35  & 9553 & 371 & 20        \\
              &     0.20     &   27.507    &  113  & 102 &  110  & 87  &  28  & 9225 & 279 & 11       \\
              &     0.50     &   30.611    &  118  & 103 &  106  & 85  &  29  & 7967 & 251 & 12       \\
              &     1.00     &   32.623    &  111  & 105 &  109  & 91  &  25  & 4372 & 190 & 8       \\
              &     2.00     &   34.654    &  110  & 98  &  104  & 87  &  26  & 3283 & 161 & 4      \\    
              &     5.00     &   36.567    &  110  & 95  &  106  & 89  &  25  & 2502 & 63 & 6       \\
              &     10.00    &   38.133    &  112  & 101 &  107  & 87  &  25  & 1607 & 49 & 2       \\              
\hline
\label{table1}
\end{tabular}
\end{table*}

\subsection{Impacts on astronomical signals }

To assess the potential risk of over-flagging genuine astronomical signals, we applied \texttt{mRAID} to data of eleven pulsars observed as part of the FAST pilot pulsar survey at intermediate Galactic latitudes~\citep{Zhi2024ApJ}. In all cases, \texttt{mRAID} effectively identified and masked RFI-contaminated frequency channels and time intervals while retaining the pulsar signals. 
Compared with the conventional \texttt{rfifind} method, the integrated S/N of the pulse profiles obtained after RFI masking are comparable (see Table~\ref{table_SR}). For example, for the bright pulsar J1832+0204\_P~\citep{Han2025RAA}, the integrated pulse profile has an S/N of 42 with \texttt{rfifind} and 44 with \texttt{mRAID}, indicating that \texttt{mRAID} does not over-flag the data.
For the nulling pulsar J1824$-$0127~\citep{Yan2024ApJ}, with a flux density of $S_{1400} = 0.59$\,mJy at 1.4\,GHz~\citep{Lorimer2006MNRAS}, we detected 84 single pulses during a 412.00\,s observation spanning 165 rotation periods using both \texttt{mRAID} and \texttt{rfifind}.
The brightest single pulse reached an S/N of 104, confirming that neither method removed genuine broadband transient signals.

\begin{table}
\centering
\caption{ Comparison of integrated S/N values for pulse profiles generated by the \texttt{rfifind} and \texttt{mRAID} RFI-mitigation techniques.}
\begin{tabular}{ccc}
\hline
\hline
Name & {S/N (\texttt{rfifind})} & {S/N (\texttt{mRAID})} \\
\hline
J1821$-$0256    &   43.4        & 49.3     \\ 
J1823$-$0154    &   66.1        & 80.0     \\ 
J1824$-$0127    &   71.9        & 72.1     \\ 
J1824$-$0132    &   124.9       & 126.8    \\ 
J1826$-$0049    &   70.3        & 68.5     \\ 
J1832+0204\_P   &   42.3        & 44.4     \\ 
J1837+0420\_P   &   16.1        & 12.6   \\ 
J1839+0543      &   12.7        & 10.2     \\ 
J1849+1001      &   39.6        & 38.9     \\ 
J1852+1200      &   18.5        & 19.1    \\ 
J1853+1303      &   319.1       & 314.0     \\ 
\hline
\label{table_SR}
\end{tabular}
\end{table}

While \texttt{mRAID} is primarily developed for time domain surveys with high time resolution, it can also be used for continuum or spectral line observations with multi-beam receiver systems.
However, our experiments with FAST 19-beam data showed that the neutral hydrogen (HI) line at 1420\,MHz was flagged as RFI by \texttt{mRAID}. This suggests that narrow-band spectral lines detectable across multiple beams, such as HI or maser emissions, can be misidentified as RFI. To address this, we implemented an option in \texttt{mRAID} that allows users to specify frequency ranges to be preserved during the masking process, ensuring that known astronomical spectral lines are retained.

\subsection{Computational performance}

We benchmarked the computational performance of \texttt{mRAID} and compared it with that of \texttt{rfifind} using a 20-core CPU cluster running on a Linux operating system.
For this evaluation, we used the same FAST data set described in previous sections. Each observation had an integration time of 412.00\,s, consisting of 8,388,608 time samples with a sampling interval of 49.152\,$\mu$s. For a subinterval duration of 1.00\,s, we computed a total of 412 CCMs as outlined in Section \ref{sec:3.2}, and performed eigenvalue decomposition for each. \texttt{mRAID} completed RFI masking for all 19 beams on a single CPU thread in 12.74\,h, while \texttt{rfifind} processed the same data with a 1.00\,s integration time in approximately 10.67\,h on the same CPU. Thus, even without parallelisation, the computational performance of \texttt{mRAID} is comparable to that of \texttt{rfifind}.
Furthermore, because the computation of CCMs and their eigenvalue decompositions are entirely independent across subintervals, the analysis of the FAST data set can, in principle, be parallelised across up to 412 CPUs for this given data set. This approach would greatly accelerate processing and achieve much more efficient RFI masking than \texttt{rfifind}.

\subsection{Key parameters for \texttt{mRAID}}

There are a few parameters that are important for RFI masking with \texttt{mRAID}. Here we describe each of them:

\begin{itemize}
    \item Bandpass normalisation parameters: as described in Section~\ref{sec:3.1}, \texttt{mRAID} employs the ArPLS algorithm for bandpass normalisation. The implementation details and performance of ArPLS have been extensively discussed in previous studies~\citep{Zeng2021MNRAS, Wang2022A&C}. Briefly, the ArPLS algorithm includes three tunable parameters: (1) \texttt{lam}, which controls the smoothness of the estimated baseline—larger values yield smoother backgrounds by imposing stronger penalties on curvature; (2) \texttt{ratio}, which sets the sensitivity of the reweighting scheme to negative deviations—smaller values make the algorithm less tolerant to downward fluctuations; and (3) \texttt{itermax}, which defines the maximum number of iterations for the asymmetric reweighted fitting process—higher values can improve convergence and baseline accuracy at the cost of increased computation time~\citep{Baek2015Ana}. Depending on the bandpass shape, frequency resolution, and bandwidth of a given observing system, these parameters can be fine-tuned to optimise the bandpass normalisation performance.
    \item Eigenvalue threshold parameter: This parameter sets the threshold for identifying RFI-affected channels. It is defined as an integer multiple of the standard deviation of the eigenvalue distribution, where the standard deviation is automatically estimated by fitting the non-RFI portion of the distribution with a Gaussian model. A smaller value results in more aggressive RFI masking. For the FAST data set used in this work, a threshold of three provided satisfactory performance, while increasing it to five resulted in only about 2\% less data being flagged as non-RFI (from about 34.6\% to 32.6\% averaged across 19 beams).
    \item Eigenvector threshold parameter: This parameter defines the threshold for identifying RFI-affected beams within a given frequency channel. It is expressed as an integer multiple of the standard deviation of the eigenvector distribution. For each beam, the standard deviation is estimated individually by fitting a Gaussian model to the eigenvector elements of non-RFI channels specific to that beam. In our experiments with the FAST data set, a threshold of one was adopted. This choice is motivated by the high sensitivity of FAST, which causes most RFI sources to be detected across multiple beams, and by the significant overlap between the eigenvector distributions of RFI-affected and clean channels. Varying the eigenvector threshold from one to three resulted in only a minor change in the fraction of masked data—from approximately 34.6\% to 31.4\% when averaged across the 19 beams, and reduced the number of beams flagged as affected by RFI by one. This indicates that, for the FAST 19-beam data set, the algorithm’s performance is only weakly sensitive to the specific choice of threshold.
    
\end{itemize}

\subsection{Limitations}

In our current implementation, bandpass normalisation is performed before computing the CCMs and carrying out the eigenvalue decomposition. For the FAST data set, our results show that \texttt{mRAID} effectively identifies both narrow-band, spiky RFI and broadband RFI. However, broadband and weak RFI may present challenges, as they can mimic bandpass variations and thus be missed by \texttt{mRAID}.

It is also worth noting that, although the current implementation of \texttt{mRAID} supports parallelisation in time, this approach becomes inefficient for short integrations due to the small number of subintervals. A potential improvement would be to enable parallelisation of CCM computation and eigenvalue decomposition across frequency channels, which are fully independent and would allow for more efficient processing.

\section{Conclusions} \label{sec:5}

This paper introduces a new RFI identification and masking software package, \texttt{mRAID}, based on the cross-correlation matrix of multi-beam data and eigenvalue decomposition. By constructing the cross-correlation matrix and decomposing it into eigenvalues and eigenvectors, the method identifies and masks RFI in different beams effectively. Compared to traditional threshold-based methods, this approach leverages the inter-beam correlation to tackle RFI under complex conditions.

Experimental results confirm the effectiveness of the proposed approach, particularly in detecting weak RFI signals and operating at short timescales. By leveraging eigenvalue decomposition, the method captures RFI characteristics through the distribution of eigenvalues and eigenvectors, enabling accurate interference identification in different beams. The use of the cross-correlation matrix further enhances robustness against both narrow-band and short-duration RFI. Compared to traditional single-beam thresholding techniques, the proposed method achieves higher detection rates while preserving unaffected data, demonstrating its superior performance in challenging observational environments.

Additionally, the computation of the cross-correlation matrix and eigenvalue decomposition is performed independently for each subinterval and frequency channel. This structure allows for efficient optimisation through parallel processing using CPU multithreading, substantially improving computational efficiency compared to RFI mitigation methods that process each beam individually. 
The \texttt{mRAID} framework is fully general and can be extended to other multibeam systems, including the PAF installed on Murriyang, the Parkes telescope. Its dependence solely on inter-beam covariance makes it adaptable to different beam layouts and instrumental responses.

\begin{acknowledgement}
We would like to thank Simon Johnston, Lister Staveley-Smith and Ron Ekers for useful discussions. This work is supported by the National Natural Science Foundation of China (Nos. 12288102, 12041304, 12273008, 12041303, 12041304, 12403046), the Major Science and Technology Program of Xinjiang Uygur Autonomous Region (No. 2022A03013-3), the National SKA Program of China (Nos. 2022SKA0130100, 2022 SKA0130104, 2020SKA0120200),  The research is partly supported by the Operation, Maintenance and Upgrading Fund for Astronomical Telescopes and Facility Instruments, budgeted from the Ministry of Finance of China (MOF) and administrated by the Chinese Academy of Sciences (CAS), the Natural Science and Technology Foundation of Guizhou Province (No. [2023]024), the Guizhou Provincial Basic Research Program (Natural Science) (QiankehejichuMS[2025]266).
This work made use of the data from FAST (Five-hundred-meter Aperture Spherical radio Telescope) (\url{https://cstr.cn/31116.02.FAST}). FAST is a Chinese national mega-science facility, operated by National Astronomical Observatories, Chinese Academy of Sciences.

\end{acknowledgement}

\paragraph{Data Availability Statement}
A Python software package of this work is available at \url{https://github.com/juntaobai/mRAID.git}.


\bibliography{sample631}


\end{document}